\documentclass[conference]{IEEEtran}
\IEEEoverridecommandlockouts
\usepackage{cite}
\usepackage{amsmath,amssymb,amsfonts}

\usepackage{amsthm}
\newtheorem{definition}{Definition}
\usepackage{graphicx}
\usepackage{textcomp}
\usepackage{xcolor}
\usepackage{algorithm}
\usepackage{framed}
\usepackage{algpseudocode}
\usepackage{url}
\def\BibTeX{{\rm B\kern-.05em{\sc i\kern-.025em b}\kern-.08em
    T\kern-.1667em\lower.7ex\hbox{E}\kern-.125emX}}

\usepackage{datetime} 
    
\begin{document}

\title{NickPay, an Auditable, Privacy-Preserving, Nickname-Based Payment System
}

\author{\IEEEauthorblockN{Guillaume Quispe${}^\dagger$}
\IEEEauthorblockA{
\textit{${}^\dagger$LTCI, Télécom Paris, IP Paris}\\
Palaiseau, France}
\and
\IEEEauthorblockN{Pierre Jouvelot${}^{\dagger, *}$}
\IEEEauthorblockA{
\textit{${}^*$Mines Paris, PSL University} \\
Paris, France}
\and
\IEEEauthorblockN{Gérard Memmi${}^\dagger$}
\IEEEauthorblockA{\textit{LTCI, Télécom Paris, IP Paris}\\
Palaiseau, France}
}

\newcommand{\GVf}[0]{\textit{GVf}}
\newcommand{\UVf}[0]{\textit{UVf}}

\maketitle

\begin{abstract}
              \noindent
        In this paper, we describe the motivation, design, security properties, and a prototype implementation of NickPay, a  new privacy-preserving yet auditable payment system built on top of the Ethereum blockchain platform. NickPay offers a strong level of privacy to participants and prevents successive payment transfers from being linked to their actual owners.
        It is providing the transparency that blockchains ensure and at the same time, preserving the possibility for a trusted authority to access sensitive information, e.g., for audit purposes or compliance with financial regulations. 
        NickPay builds upon the Nicknames for Group Signatures (NGS) scheme, a new signing system based on dynamic ``nicknames'' for signers that extends the schemes of group signatures and signatures with flexible public keys. 
        NGS enables identified group members to expose their flexible public keys,  thus 
     allowing direct and natural applications such as auditable private payment systems, NickPay being a blockchain-based prototype of these.

\end{abstract}


    \section{Introduction}
    \label{sec:introduction}

    Information-technology-based payment systems managed by financial institutions  require their clients to trust them to operate with integrity while also keeping their data private.
    In 2008, the introduction of Bitcoin~\cite{nakamoto2008bitcoin} enabled such computational integrity to no longer rely on a single entity but rather on a permissionless distributed set of validators.
    This technology, often named Distributed Ledger Technology (DLT), is of great interest to the financial world\footnote{https://www.federalreserve.gov/econres/notes/feds-notes/governance-of-permissionless-blockchain-networks-20240209.html}.

    However, new problems arise, particularly since it requires the public verification of transactions involving user identities hidden under pseudonyms.
    If bank statements are usually kept private from the public, all user activity on DLT is public and, without the addition of a privacy feature, is only protected by pseudonyms.
    If the link between a pseudonym and its user is leaked for a given transaction, inadvertently or by algorithmic analysis \cite{reid2013analysis}\cite{meiklejohn2013fistful}\cite{fleder2015bitcoin}, this bank statement becomes publicly available.
    Such a breach of privacy constitutes a critical problem for users as for regulation authorities, especially now that new user-protection legislation has been introduced~\cite{gdpr2016general,INTprivacy24}.
    Obviously, proposals to address blockchain-related privacy issues came early, e.g., via the introduction of zero-knowledge proofs to blockchains~\cite{sasson2014zerocash}, or the combination of ring signature and stealth address technologies~\cite{noether2016ring}.
    
    But privacy is not secrecy:
    ``A private matter is something one doesn’t want the whole world to know, but a secret matter is something one doesn’t want anybody to know''~\cite{hughes1997cypherpunk}.
    Secrecy can protect behaviors seen as illegitimate, e.g., money laundering, and can threaten the interests of others.
    Solutions offering secrecy~\cite{noether2016ring} still allow illegal activities, and in 2018, it was estimated~\cite{moser2017empirical} that close to 25\% of Monero-based transactions (from~\cite{noether2016ring}) were illicit.

Since current traditional financial agents aim to integrate more and more  DLT into their portfolio of offered services, it is important to propose flexible solutions that facilitate its integration into the current financial environment.
    In particular, finding a proper equilibrium between privacy and regulation requirements is of paramount importance.
    We advocate here for ``privacy by design'', as it should not be optional, and  malicious behavior should  be prevented or sanctioned by regulatory-compliant means, e.g., Know Your Customer (KYC) checks, auditability and accountability~\cite{chatzigiannis2021sok}.

\paragraph{Overview}
   We describe in this paper the design, security properties and implementation details of NickPay, a  new privacy-preserving yet auditable payment system built on top of the Ethereum blockchain platform~\cite{Eth14}. It offers participants identity privacy and prevents successive payment transfers to be linked to their actual owners, all this while providing the transparency that blockchains ensure. 
   
   At the same time, NickPay preserves the possibility for a trusted authority to access  sensitive payment information, e.g., for audit purposes or compliance with financial regulations. 
    The auditability property NGS offers operates at two levels, internal and external.
    The internal auditing is similar to the global supervision of \cite{chen2020pgc} where a supervisor can audit the system without interacting with the auditees, i.e., the users.
    But such a system can also be audited by an external entity. 
    In this case, the supervisor should be able to answer any query from the external auditor to identify potential malicious transactions and members, and expose those to him selectively.
        
        NickPay builds upon the Nicknames for Group Signatures (NGS) scheme, a recently introduced signing system based on dynamic ``nicknames'' for participants that extends the existing schemes of group signatures (GS) and signatures with flexible public keys (SFPK) (see Section~\ref{sec:background}, for background information). 
        NGS enables selected participants, identified by a dedicated ``issuer'' as members of a ``group'', to expose their flexible public keys, named ``master public keys'', thus 
     allowing direct and natural applications such as auditable private payment systems.

    To illustrate how NickPay works on a simple case, assume that, after registration, Alice wants to send tokens (money, in Ethereum's speak) to Bob. She then derives a nickname $nk$ from Bob's master public key $mpk$ and signs the Ethereum transaction sending her tokens to $nk$.
    The smart contract handling transfer settlements on the blockchain has to do two things before updating the balances accordingly.
    First, it verifies whether Bob's nickname $nk$ is valid, i.e., passes the group membership verification step \textit{\GVf}, and then that the sender Alice owns the sender nickname mentioned in the transaction.
    The NGS traceability property (see Section~\ref{sec:ngs}) prevents at any time a user not in the group managed by the contract to pass the \textit{\GVf} algorithm to spend money.
    Also, NGS allows a ``supervisor'', i.e., the NGS so-called ``opener'', at any time to know the balance of each member.
    Bob can then determine if $nk$ is in his equivalence class of nicknames thanks to an NGS-provided ``trapdoor'' and prove ownership of $nk$, thus leading to the completion of the money transfer.
    If need be, the supervisor can iterate over all the user (encrypted) trapdoors to identify the member owning $nk$, here Bob, and prove it to an external auditor.

    \paragraph{Contributions}
    We present the following new results:
    \begin{itemize}
   
        \item the design  of NickPay, a DLT-based, anonymous and auditable payment system based on the NGS signature scheme for anonymity and auditability;
        \item the specification of security properties, some of which new in the security domain, provided by NickPay, directly inherited from NGS;
        
        \item a prototype implementation of NickPay on top of the Ethereum blockchain, with preliminary performance data;
        \item an implementation in Solidity of the NGS verification functions, inspired by the existing NGS Rust code~\cite{NGS}.
    \end{itemize}

    \paragraph{Organization}
    After this introduction, in Section~\ref{sec:introduction}, Section~\ref{sec:background} presents some background information that helps grasping the main features of NGS, introduced in  Section~\ref{sec:ngs}, and in particular its security model, described in Section~\ref{sec:security}.
    In Section~\ref{sec:application}, we present the design, properties and Ethereum-based implementation of NickPay, the application of NGS to an anonymous yet auditable payment system. Section~\ref{sec:related} covers the related work. We conclude and discuss future work in Section~\ref{sec:conclusion}.

    \section{Background}\label{sec:background}
    This section presents the basic signature-related concepts that are at the core of NickPay. We use, however, a dedicated section to describe in detail NGS, the specific signature scheme it uses (see Section~\ref{sec:ngs}).

    \paragraph{Digital signature}
        A {\em digital signature scheme} $DS=\{\textit{KeyGen},\ \textit{Sig},\ \textit{Vf}\}$ consists of 3 algorithms:
        \begin{itemize}
            \item $\textit{KeyGen}: 1^\lambda\rightarrow (sk, pk)$, a key generation algorithm that outputs a pair of secret and public keys $(sk,pk)$ under security parameter $\lambda$;
            \item $\textit{Sig}: (sk,m) \rightarrow \sigma $, a signing algorithm that takes a secret key $sk$ and a message $m \in \{0,1\}^*$ and outputs the DS signature $\sigma$ for $m$;
            \item $\textit{Vf}: (pk,m,\sigma) \rightarrow \{0,1\}$, a verification algorithm that takes a signature $\sigma$, a message $m$ and a public key $pk$ and outputs 1, if $(m,\sigma)$ is valid (i.e., the secret key used to build $\sigma$ and $pk$ match under DS), and 0, otherwise.
        \end{itemize}

Typically, a user will first call $KeyGen$ to get his own signing keys. Then, whenever he wants to send a message $m$ carrying his seal of approval $\sigma$, he will use the $Sig$ function, the output of which is guaranteed to pass the \textit{Vf} test function.
    Note that, if only the owner of the secret key $sk$ can sign a message $m$ that can be properly verified, the verification can be performed by anyone having access to $pk$ and $m$. 
    
    Signatures are designed so that matched $(m, \sigma)$ pairs cannot be forged. The standard security property for a digital signature scheme is ``existential unforgeability
    under chosen message attacks'' (EUF-CMA) \cite{goldwasser1988digital}.
    Informally, it states that, given access to a signing oracle, it is computationally hard (in terms of $\lambda$) to output a valid pair ($m$, $\sigma$) for a message $m$ never before submitted to the signing oracle.

    \paragraph{Group signature} A \textit{group signature scheme} (GS) is a DS scheme introduced by D. Chaum and E. van Heyst in \cite{chaum1991group} circa 1991.
    Such a scheme provides a relative anonymity to the signers (called ``group members'') as the signature reveals to the public the sole name of his group, and only group members can sign messages.
    But, it also offers an ``opening'' feature, since a ``group manager''~\cite{getshorty} can open signatures, i.e., identify a signer based on his signature and thus break anonymity (usually for himself or other stakeholders such as an auditing authority).
    A \textit{dynamic group signature scheme} (DGS) \cite{bellare2005foundations}
    extends GS so that the group manager has the ability to accept users to join the group at any moment,  or perform revocation, as in \cite{Bre01,LiPe12}. 

     (D)GS are endowed with formal security properties defined via oracle-equipped adversaries and probabilistic reasoning:   
     \begin{itemize}
         \item
     ``Correctness'' ensures that a signature from an honest member can be verified successfully with probability 1;
     \item  ``Anonymity'' ensures (i.e., with a large probability) that no adversary can identify the signer of a target group signature;
     \item  ``Non-frameability'' ensures that an honest user cannot be falsely accused of having signed a message;
     \item  ``Traceability'' ensures that no user can produce a valid group signature that is not traceable by the opener;
     \item ``Opening soundness'' ensures that no adversary can produce a signature that can be opened to two distinct users.
    \end{itemize}
    Note that many other security properties such as unforgeability were proposed before, but Bellare, Shi and Zhang~\cite{bellare2005foundations} showed that their properties, sketched above, encompass these.

 \paragraph{Stealth signature}

    The notion of \textit{stealth address}, or \textit{stealth signature} (SS), introduced in 2014 in the Bitcoin blockchain~\cite{courtois2017stealth}, can be used in crypto-currency systems. There, in a typical transaction, the sender would send some tokens (the amount of which could also be hidden) to $pk'$, a one-time, dynamically derived public key of the recipient's original public key $pk$, and the latter could unlock them by taking advantage of his (unique) secret key. This mechanism increases the privacy level of transactions, but can lead to the augmented use of such infrastructures for law-breaking financial operations.

    \paragraph{Signature with flexible public key} A \textit{signature with flexible public key scheme} (SFPK)~\cite{backes2018signatures}  partitions the key space of some DS into equivalence classes induced by some relation $\mathcal{R}$.
    In other words, one can transform a signer's public key $pk$ into a different representative $pk'$ of the same $\mathcal{R}$-equivalence class without the help of $pk$'s secret key: the old and new public keys are thus related by $\mathcal{R}(pk, pk')$.
    Additionally, the SFPK class-hiding property states that, without a so-called ``trapdoor'', it should not be possible to determine whether two keys belong to the same class.
    Implicitly, owning such a trapdoor would imply user traceability as the signer could determine whether a public key belongs to his equivalence class and, thus, recover its signing key.
    Hence, the primitive of key transformation via equivalence classes offers a way to communicate anonymously, as demonstrated in multi-user settings~\cite{backes2018signatures}, SPFK being a possible implementation of SS.

    \section{Nicknames for Group Signatures}\label{sec:ngs}

    NickPay is build on top of a new, recently introduced DGS scheme called Nicknames for Group Signatures (NGS)~\cite{NGS}. NGS merges concepts present in (D)GS  and SFPK in order to provide both anonymity and auditability within a communication system.
    From (D)GS, it inherits its group-management policy, where only users interested in joining the group and accepted by the group manager (here called ``issuer'') can perform  operations in a publicly verifiable way.
    It also provides a private communication feature inspired by SFPK, where members are publicly identified by an arbitrary representative of their equivalence class that we call here ``master public key''.
    
    In NGS, anyone can derive a key, called a ``nickname'', for a group member from his master public key. More precisely, during the joining phase of the NGS protocol by user $i$, the issuer associates an equivalence class $[{mpk}]_\mathcal{R}$ to $i$ and publishes $i$'s master public key, $mpk$.
    NGS then allows anyone to create a ``nickname'', i.e., a different representative ${nk}$ of the class $[{mpk}]_\mathcal{R}$, without access to the secret key controlling ${mpk}$. 
    This nickname can then be used to privately refer to $i$ when needed. 
    
    Similarly to both GS and SFPK, NGS provides some anonymity, in that it hides the link between members and their nicknames, i.e., neither nicknames nor  operations referring to different nicknames can be linked together, even when  actually dealing with the same underlying user.
   However, NGS also enjoys the user traceability property of SFPK, which allows a group member $i$ to retrieve his nicknames in $[{mpk}]_\mathcal{R}$ via a trapdoor, leading to a weakened version of anonymity, called ``selfless anonymity''~\cite{boneh2004group}.
   
    Finally, NGS provides a GS-opening-like feature, where an opener is able to provably identify a member from any of its nicknames, thus offering auditablity assurance for authorized participants.
NGS can thus be viewed as an extension of GS with communication capability between users, {\em à la} SFPK.


    \subsection{Interface}

    NGS is abstracted over roles, types, variables, and functions that are instantiated when defining a precise implementation.

    \begin{definition}[NGS Roles]
    In the NGS scheme, each participant, or User, has a role that provides him with the rights to handle particular data, global variables or abilities of the NGS scheme. We briefly describe the five key NGS roles.
        \begin{itemize}
            \item Verifier: any user or participant can act as a Verifier, to check that a given nickname does exist. A user can also choose not to join the group but still to interact with its members, for instance, to verify that a message is signed with an existing nickname or to produce nicknames.
            \item Issuer: the Issuer is the only role providing the ability to authorize users to become group members. 
            \item User: a user becomes a {Group Member} by having his ``join request'' accepted by the issuer. Then, he will be able to sign messages. 
            \item Opener: the Opener is the only role providing the ability to ``open'' a nickname, i.e., to unveil the underlying group member's identity, together with a proof of it. 
            \item Judge: the Judge role can be adopted to check that a proof supposedly linking a user to a nickname (following the opening of a nickname)  is valid. 
        \end{itemize}

    \end{definition}

     Most of the time, one would also define the additional role of {Group Manager}, which would endorse both roles of issuer and opener. 

    \begin{definition}[NGS Types]
        NGS uses the following set of generic types.
        \begin{itemize}

            \item A Registration information is a structure, usually named $reg$, that
            contains the necessary identification elements of a user, saved in the registration table $\textbf{reg}$ defined below.
            The contents of $reg$ is implementation-dependent, but can include, for example, the encryption of the users' trapdoors $\tau$.

            \item A Join request is a structure, named $reqU$, produced by a user requesting to join the group. $reqU$ contains the necessary implementation-dependent elements for a joining request to be handled by the issuer. It can be seen as a synchronization element between the user and the issuer during the group-joining algorithm.


        \end{itemize}
    \end{definition}

    \begin{definition}[NGS Global Variables]
        NGS is build on the following set of global variables.
        \begin{itemize}
            \item $DS=\{\textit{KeyGen},\textit{Sig},\textit{Vf}\}$ is a digital signature scheme that will be used in the implementation.  It is assumed that its public keys are certified by a certification authority (CA).
            \item $\textbf{reg}$ is the user-indexed registration table controlled by the issuer, who has read and write access to it.
            The opener is given a read access to it also.
            \item $\textbf{mpk}$ is the user-indexed master public key table, publicly available and used to define nicknames.
            \item $\textbf{upk}$ is the publicly available table of users' public keys for $DS$ scheme.
        \end{itemize}
    \end{definition}
    \begin{definition}[NGS Scheme]
        A {\em nicknames for group signatures scheme} (NGS) is a tuple of functions, each one  related to one key role in NGS-based protocols, (User, Group Member, Issuer, Opener, Verifier, Judge), defined as follows. 

        \begin{itemize}
            \item $IKg: 1^\lambda \rightarrow (isk,ipk)$ is the key generation algorithm that takes
            a security parameter $\lambda$ and outputs an issuer $DS
            $ secret/public keys ($isk,ipk$).
            \item $OKg: 1^\lambda \rightarrow (osk,opk)$ is the key generation algorithm
            that takes a security parameter $\lambda$ and outputs an opener's $DS$ secret/public keys ($osk,opk$).
            \item $UKg: 1^{\lambda} \rightarrow (usk,upk)$ is the user key generation algorithm that produces $DS$ secret/public keys $(usk,upk)$, its second component being stored in $\textbf{upk}$. 
            \item $Join: (usk,ipk,opk) \rightarrow (msk,\tau, reqU)$, the user part of the group-joining algorithm, takes a user's secret key $usk$ and the issuer and opener public keys and outputs a master secret key $msk$ along with a trapdoor $\tau$. $reqU$ will then contain information necessary to the Issuer to run the second part of the group-joining algorithm.
            \item  $Iss: (i,isk,reqU,opk) \rightarrow ()$ or $\perp$, the issuer part of the group-joining algorithm, takes a user $i$, an issuer secret key $isk$, a join request $reqU$ and the opener public key $opk$,  updates the registration information $\textbf{reg}[i]$ and master public key  $\textbf{mpk}[i]$, and creates an equivalence class for the user's nicknames. it returns $\perp$, in case of rejected registration.
            \item $Nick: mpk \rightarrow nk$ is the nickname generation algorithm that takes a master public key $mpk$ and creates a nickname $nk$ belonging to $[mpk]_\mathcal{R}$.
            \item $Trace: (\tau,nk) \rightarrow b$ takes as input the trapdoor $\tau$ for some equivalence class $[mpk]_\mathcal{R}$ and a nick $nk$ and outputs the boolean $b=(nk \in [mpk]_\mathcal{R})$.
            \item $Sig: (nk,msk,m) \rightarrow \sigma$ takes a nickname $nk$, a master secret key $msk$ and the message $m$ to sign and outputs the signature $\sigma$.
            \item $\textit{\GVf}~: (ipk,nk) \rightarrow b$ is the issuer verification algorithm that takes a issuer public key $ipk$ and a nickname $nk$ and outputs a boolean $b$ stating whether $nk$ corresponds to a user member of the group or not.
            \item $\textit{\UVf} : (nk,m,\sigma) \rightarrow b$ is the user verification algorithm, taking as input a nickname $nk$, a message $m$ and a signature $\sigma$ of $m$ and outputting a boolean stating whether $\sigma$ is valid for $m$ and $nk$ or not.
            \item $Open: (osk,nk) \rightarrow (i,\Pi) \text{ or} \perp$,  the opening algorithm, takes an opener secret key $osk$ and a nickname $nk$.
            It outputs a user  $i$ with a proof $\Pi$ that $nk$ is a member of $i$'s equivalence class,  or $\perp$, if no such user exists. 
            \item $Judge: (nk,ipk,i,\Pi) \rightarrow b$ is the judging algorithm that takes a nickname $nk$, an issuer public key $ipk$, 
            a user $i$ and an opener's proof $\Pi$ to be verified and, using the user public key  $\textbf{upk}[i]$,  outputs a boolean $b$ stating whether the judge accepts the proof or not.
        \end{itemize}
    \end{definition}

    \subsection{Security model}\label{sec:security}

   DGS have been studied for several decades and are well formalized~\cite{bellare2005foundations}\cite{bellare2003foundations}. Their security properties (see Section~\ref{sec:background}) are based on thought experiments based on oracles and adversaries equipped with varying levels of abilities.
 The NGS security model is heavily dependent on such foundations, with some necessary modifications, which are highlighted here,
    \begin{itemize}
        \item 
  The group verification function of GS is split into the NGS $\GVf$ and $\UVf$ functions for better modularity, as illustrated in the NickPay operations (see Section~\ref{sec:application}).
    A forgery in NGS would consist of an improperly produced nickname $nk$ passing $\GVf$, along with a corresponding signature $\sigma$ passing $\UVf$ on $nk$.
    NGS non-frameability and traceability properties thus require, in their definitions, an adversary able to output such a forgery.
    \item
    The NGS opening function takes, in addition of the opener secret key, only nicknames as arguments, not  additional signatures $\sigma$, as is done in DGS.
    In practice, this means that, when a nickname $nk$ is opened, either it returns the identity of the user controlling it, or it fails.
    But in the latter case, the opener can be assured that $nk$ will not come with a valid $\sigma$; otherwise it would break traceability.
    However, in the NGS anonymity experiment used to define this property, the adversary is restricted and close to a Chosen-Plaintext Attack (CPA), with no access to a ${Trace}$ oracle.


\end{itemize}

    \section{NickPay}\label{sec:application}

    In this section, we describe our main contribution: NickPay, an NGS-based, auditable, and anonymous distributed (i.e., ledger-based) payment system, a prototype of which is implemented on top of the Ethereum blockchain platform. In addition to the NGS scheme, this system introduces two new minting and transfer request functions, together with their corresponding mint and transfer settlements.

\subsection{Presentation}
    Roles in NickPay are close to the ones given in NGS.
    We keep the Issuer, Verifier, and User denominations, while the NGS Opener and Judge are renamed Supervisor and Auditor,  respectively.
    We introduce the notion of ``{minting authority}'', an entity able to create money and not necessarily in the group.
    
    An account $(nk,\beta,\eta)$ is of a particular type consisting of $nk$, a nickname identifying the account owner, $\beta$, the account balance, and $\eta$, a so-called nonce for this account.
    In practice, a nonce is a counter incremented each time a signature on this account's nickname $nk$ is verified, and a payment transaction is settled; it is intended to be used to prevent a malicious user from sending twice the same transaction.
    We introduce the account public array $\mathbf{acc}$, mapping nicknames to their account.
    
    A payment $tx$ is a transaction between two accounts; it is  defined as a tuple $(nk_s,nk_r,v,\eta)$ where $nk_s$ is a sender nickname, $nk_r$, a receiver nickname, $v$, a number of tokens (to model any arbitrary monetary unit), and $\eta$, the nonce in $nk_s$'s account, which gets incremented once the payment is completed.
    We suppose given a hash function $H$, used notably to compute digests of transactions, to be signed and verified.
    
    Using platform-specific ``contracts'', the ledger manages and exposes the key registry containing the master public keys $\textbf{mpk}$ and maintains also $\textbf{acc}$. It makes two functions available: $\textit{Mint}$, which creates tokens for a specified recipient, and $\textit{Transfer}$, which exchanges some tokens and, additionally, checks that the sender has enough of those. 

\subsection{Operations}
    We describe below the main operations that characterize NickPay, covering a typical use case.

    \paragraph{Setup} 
         The issuer and supervisor run $IKg$ and $OKg$, respectively, to obtain their pair of secret/public keys $(isk,ipk)$ and $(osk,opk)$, respectively.
    The issuer will decide (or not, e.g., via a dedicated KYC  process) to grant users access to the group it manages, in which case the supervisor will be able to track, i.e., open, their nicknames used in messages' signatures.
    
    In addition, a verifier contract is instantiated  with the issuer public key $ipk$.
    The master public key and token contracts are also instantiated.
    The token contract instance directly uses the verifier's one.

\begin{figure}
        \centering
        \includegraphics[width=0.55\linewidth]{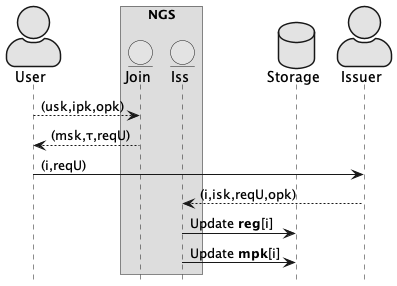}
        \caption{Join synchronization process: this sequence diagram illustrates how a user $i$ interacts with the NGS library and the issuer to join (request accepted, in this case) the group.}\label{fig:ngs-group-synchro}
    \end{figure}

    \paragraph{Group-joining synchronization process}
    Prior to joining the group, the user $i$ must run $UKg$ to obtain his secret/public keys $(usk,upk)$.
    Then, to effectively enable the user $i$ to join the issuer's group, the following three 
    steps must be performed (see~Figure~\ref{fig:ngs-group-synchro}).

    (1) The user $i$ must run the $Join$ function, providing him notably a join request (to be submitted to the issuer) and his putative secret group-signing key $msk$.  
    
    (2) Then, the issuer must run the $Iss$ function to verify the correctness of this join request, in particular with respect to the user's public key, $upk$, and the supervisor public key, $opk$.
    The issuer creates the user $i$'s master public key, $mpk$, used to create his equivalence class and associated nicknames, and adds it to the public $\textbf{mpk}$ table.
    
    (3) Finally, after checking that the issuer information is correct, i.e., $\GVf(ipk,mpk)$, the user can locally 
    store the now-accepted $msk$.

\begin{figure}
        \centering
        \includegraphics[width=0.7\linewidth]{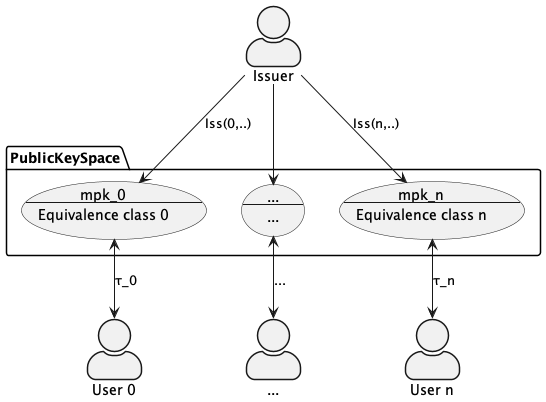}
        \caption{Join protocol: the issuer grants the user $i$ an equivalence class by defining $\textbf{mpk}[i] = mpk_i$.
        On his side, the user $i$ can identify any public key in his equivalence class by using his trapdoor $\tau_i$.
        Note that $\textbf{mpk}[i]$ is itself in the equivalence class of user $i$, as it is just an arbitrary element inside it. }
        \label{fig:ngs-join}
    \end{figure}

    If the protocol succeeds, user $i$ obtains thus a confirmation that his master secret key $msk$ is valid for the equivalence class $[mpk]_\mathcal{R}$, allowing him to sign on behalf of the group, along with his trapdoor $\tau$, while its encrypted version is stored in $\textbf{reg}$, to allow the supervisor to identify his nicknames in the future (see Figure~\ref{fig:ngs-join}).

    \paragraph{\textit{Mint} request} Suppose the minting authority wants to mint $v$ tokens for user $i$.
    It does so by first computing a nickname $nk = Nick(\mathbf{mpk}[i])$ of user $i$.
    The minting authority then creates a mint-specific  transaction $tx=(nk,v,\eta)$, requesting to mint $v$ tokens to $nk$, and the blockchain signature (e.g., via ECDSA, in Ethereum) $\Tilde{\sigma}$ of  $m=H(tx)$. 
    It then sends $(tx,\Tilde{\sigma})$ to the $\textit{Mint}$ function hosted by the blockchain.

    \paragraph{\textit{Mint} settlement} The blockchain \textit{Mint} function handles signed mint-specific transactions $(tx,\Tilde{\sigma})$, with $tx = (nk,v,\eta)$.
    \begin{itemize}
        \item The request is passed to the verifier, who first checks whether the recipient nickname $nk$ is part of the group by running $\GVf(ipk, nk)$ with the issuer public key $ipk$.
        It then checks the validity of the minting authority's blockchain signature $\Tilde{\sigma}$.
        \item Then, $v$ tokens are added to $\mathbf{acc}[nk]$'s balance.
    \end{itemize}

   \begin{figure}[H]
        \centering
        \includegraphics[width=0.9\linewidth]{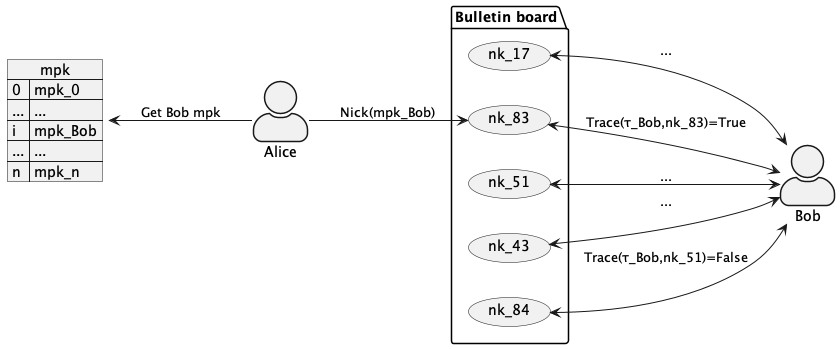}
        \caption{Nick protocol: Alice can create, using the $\textit{Nick}$ algorithm on Bob's $mpk$, a new nickname, e.g., $nk_{83}$, for Bob  and use it to post some signed message on the blockchain. 
        Bob can then check if he can unlock the last posted message, here addressed to nickname $nk_{83}$, with the $\textit{Trace}$ function.
        }\label{fig:ngs-nick}
    \end{figure}
    


  
    \paragraph{Tracing} Over time, user $i$ can scan the nicknames recorded in $\textbf{acc}$ by calling the $Trace$ function with his trapdoor $\tau$ to filter the ones he owns and can therefore unlock.
    The value $Trace(\tau, nk)$, with $nk$ the account nickname in $\mathbf{acc}[k]$ for some user $k$, is true whenever $nk \in [\mathbf{mpk}[i]]_\mathcal{R}$. This function can be used as a subroutine of the signing protocol (see Figure~\ref{fig:ngs-nick}).


   \begin{figure}
        \centering
        \includegraphics[width=0.8\linewidth]{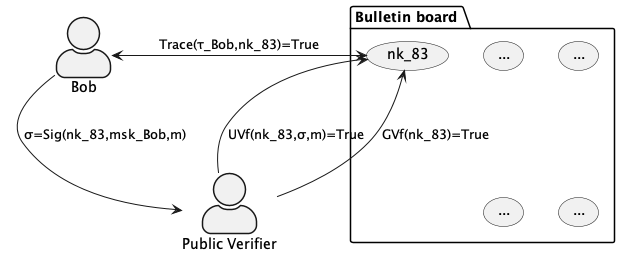}
        \caption{Sign protocol: Bob can  sign (using a signature $\sigma$) a message $m$ of his choice, proving he controls the nickname $nk_{83}.$
        The verifier can then check that this nickname belongs to the group and the validity of $\sigma$.
        }\label{fig:ngs-verif}
    \end{figure}

    \paragraph{\textit{Transfer} request}
    Assume that user $i$, in possession of $v$ tokens on the account linked to one of his nicknames, $nk$, of nonce $\eta$, wants to transfer $t\leq v$ tokens to user $k$.
    He creates a nickname $nk' = Nick(\mathbf{mpk}[k])$ for user $k$.
    He then signs the transaction $tx=(nk,nk',t, \eta)$ transferring $t$ tokens to $nk'$,  and obtains $\sigma = Sig(nk, msk,m)$, with $msk$ being the private master key of user $i$ and $m=H(tx)$.
    He then sends $(tx,\sigma)$ to the $\textit{Transfer}$ function hosted on the blockchain.

    \paragraph{\textit{Transfer} settlement} The blockchain \textit{Transfer} function handles the pairs $(tx,\sigma)$, with $tx = (nk,nk',v,\eta)$, provided as a transfer request in two steps.
    \begin{itemize}
        \item As in the mint case, the $\GVf$ function verifies that $nk'$ is in the group. Next, $\UVf$ verifies that the signature $\sigma$ is correct, meaning that the group member controlling $nk$ consents to this transfer. 
        \item Then, if $\eta$ has been seen before or in case of insufficient funds in the account $a = \mathbf{acc}[nk]$, the verifier aborts. Otherwise,  $a$ is updated by subtracting $v$ tokens to its balance, and  the account $a' = \mathbf{acc}[nk']$, by adding $v$ to it.
        Finally, the nonce of $a$ is incremented.
    \end{itemize}

       \begin{figure}
        \centering
        \includegraphics[width=0.6\linewidth]{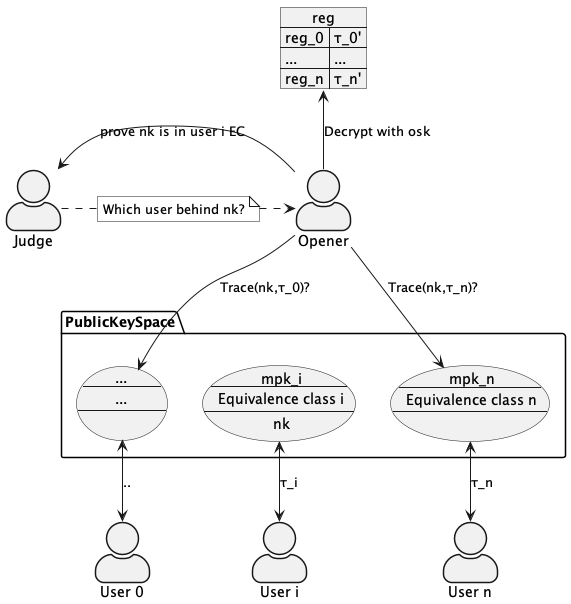}
        \caption{Open protocol: The supervisor decrypts users' trapdoors in $\textbf{reg}$ and try each trapdoor $\tau$ to open the nickname $nk$ asked by the auditor. }
        \label{fig:ngs-open}
    \end{figure}

    \paragraph{Opening} For any nickname $nk$, the supervisor can identify which user, $i$,  controls it by decrypting one-by-one the users' trapdoors stored in $\textbf{reg}$ (see Figure~\ref{fig:ngs-open}).
    He produces a publicly verifiable proof $\Pi$ stating that $i$ indeed controls $nk$.
    

    \paragraph{Audit}
    An  auditor suspecting a possible malicious operation on a nickname $nk$ for a specific transaction can ask the supervisor to reveal the actual identity behind $nk$.
    The supervisor would then call $Open$, and send the resulting proof $\Pi$ stating that a certain user, $i$, controls $nk$ to the  auditor.
    The latter can verify the validity of $\Pi$ by calling the $Judge$ function, thus being assured that $i$ indeed controls $nk$.

    \subsection{Properties}



As in \cite{chen2020pgc}, we consider here the NickPay properties defined at the transaction layer, not the network level.

\paragraph{Authenticity} 
    A Decentralized Anonymous Payment (DAP) system such as NickPay requires authenticity.
    Authenticity ensures that, given any honest user $i$, no adversary can steal his tokens, i.e., tokens locked in all nicknames $nk$ such that $Trace(\tau, nk)$, with $\tau$ the trapdoor of $i$.
For NickPay, DAP authenticity is ensured by the non-frameability of NGS~\cite{NGS}.

Note that NGS traceability ensures that tokens tagged with a certain $nk$ that passes the $\GVf$ group verification function and cannot be identified, i.e., $Open(osk, nk) = \perp$, will not be spent, i.e., are lost.

\paragraph{Anonymity}

To deal with anonymity in NickPay, we need to introduce the new notion of ``honest sender anonymity'', stating that no adversary can identify a user from a nickname generated for him by an honest sender. In NickPay, anonymity is ensured by NGS's corresponding property~\cite{NGS}.

    \paragraph{Auditability} NickPay offers two types of auditability, internal and external.
    \begin{itemize}
       \item  Internal auditing is similar to the global supervision mentioned in \cite{chen2020pgc}, where a supervisor gets extra privileges for this task.
    It is non-interactive in NickPay, since the supervisor does not require the auditee's, i.e., in this case, the user's, consent.
    Typically, an internal audit process tries to ensure that laws are respected and that external audits would not sanction any activity.

    \item External auditing is an interactive protocol with the supervisor.
    In a typical case, the auditor, who suspects some malicious behaviour from a certain user nicknamed $nk$, can ask the supervisor to reveal the user controlling $nk$.
    Additionally, this auditor could ask for the ability to perform the supervision~\cite{chen2020pgc} for a certain user $i$; the supervisor would then provide $i$'s trapdoor.
    \end{itemize}
    
    For a DAP that intends to offer auditability features, it is pertinent to address the property of ``secure auditing''~\cite{chen2020pgc}.
    This property, added to anonymity, if satisfied, would ensure, first, that internal auditing is robust, i.e., that the supervisor cannot break the authenticity of the DAP.
    Secondly, it ensures that external auditing is sound, i.e., that an external auditor will reject false claims from the supervisor and accept correct ones, all this with minimal information disclosure.

    For NickPay, DAP auditing robustness is ensured by the non-frameability of the NGS scheme~\cite{NGS}, and 
auditing soundness, by NGS opening-soundness and traceability properties. The minimal information-disclosure constraint derives from the zero-knowledge nature of the proofs handled within NGS.


    \subsection{Implementation architecture}
    In this section we provide more details about the prototype implementation of NickPay we built on the Ethereum blockchain. We used Solidity for smart contracts, and Rust for off-chain computations. 
  
    The key registry exposing $\textbf{mpk}$ is managed in a dedicated smart contract controlled by the issuer, thus capable of adding any master public key to the group. Anyone accessing the blockchain can read the registered master public keys.  
    
    The $\textit{Mint}$ and $\textit{Transfer}$ functions described in the previous section are exposed in an ERC-20 token contract, the Ethereum's token standard. This token manages the balances of the nicknames' accounts in a hashmap that takes  addresses of  nicknames as keys and their balances as values. Addresses are derived from nicknames by computing the \textit{keccak256} hash function on the concatenated elements of nicknames. 

    The verifier functions $\GVf$ and $\UVf$ are exposed in a library that is used by a verifier contract for the particular group, i.e., for the $ipk$ of the issuer chosen for this prototype. 

    Only the minting authority can call the \textit{Mint} function to create tokens for a specified nickname. This function receives the amount of tokens and recipient nickname, executes the $\GVf$ function of the NGS Solidity library to verify that the recipient is in the group and updates the balance of the recipient, if it is.
    Lastly, a specific event, called ``{Announcement}'', is emitted, exposing the recipient nickname. Note that nicknames are not stored in a contract for efficiency purpose.
    
     For the \textit{Transfer} function, we use the ERC-2771 standard for meta-transactions. Indeed, the NickPay accounts in $\textbf{acc}$ are not directly compatible with Ethereum accounts and do not hold ether, the coin necessary to send transactions and pay gas. To address this issue, we programmed and deployed a standard-specific so-called TrustedForwarder contract as the entry point to interact with any application involving nicknames, including the NickPay token. It manages a hashmap from addresses of  nicknames to nonces. 
     
     In this prototype, the user behind a certain $nk$ first signs off-chain an  EIP-712-compliant transfer message with the NGS signing algorihtm.  EIP-712 standardizes the hashing and signing of type-structured data and is used in the ERC-2771 standard. The signed message is then handled by a relayer, a specific blockchain agent that packages it into a so-called meta-transaction before signing  and sending it to the TrustedForwarder contract. Any Ethereum account with gas can relay those transactions to the publicly available $\textit{execute}$ function of the latter contract.  
     This function receives the meta-transaction, calls the $\UVf$ function of the verifier contract to validate the NGS signature and updates the nonce of the sender's nickname address. It then forwards the request to the $\textit{Transfer}$ function, which verifies that the recipient is in the group, with $\GVf$, and emits an Announcement as in the minting function; the sender's and receiver's balances are then updated. Note that the nickname-tracing and opening functions are executed off-chain, after parsing the Announcement events to check the recipient nickname.

     We use the Arkworks library for the cryptographic operations, including  pairings, on the BN254 elliptic curve  that underlies the NGS instantiation. We used the Rust-based generic NGS library that exposes the interface of the NGS scheme described in Section~\ref{sec:application}~\cite{NGS}. On top of this NGS library, we finally build NickPay, in particular the Solidity contracts, as specified  above.

\subsection{Performance}

Even though this implementation of NickPay is more a proof-of-concept attempt than a production-ready complete payment solution (which is left as future work), some elements of run-time performance are provided here.
\begin{table}
\begin{center}
\begin{tabular}{ |c||c|c|c|  }
 \hline
 & \textit{Mint}&\textit{Transfer}\\
\hline
 Usage (gas)& 222,2&258,8  \\
 \hline
Gas price (Gwei)&3.591&3.791 \\
 \hline
 Price ($10^{-3}$ ETH) &0.798&0.981\\
 \hline
Price (USD)&2,64&3,20 \\
\hline
\end{tabular}  
\end{center}
\caption{NickPay resource usage (per Ethereum function call). Gas prices differ, since these functions were run at different times of the day (Dec. 22, 2024).}
\label{tbl:gas}
\end{table}
Table~\ref{tbl:gas} displays the amount of gas and (current) price, in ether and USD, needed to run, on the Sepolia testnet, the Ethereum  on-chain \textit{Mint} and \textit{Transfer} NickPay functions. The gas usage is the most stable and informative data; the gas price, which corresponds to the unitary price of gas in Gwei, i.e., $10^{-9}$ ether (ETH), the native crypto-currency of Ethereum, fluctuates during the day. The ETH fees are computed by multiplying these two values, and the USD prices are obtained by multiplying the ETH fees by the current exchange rate (as of Dec. 22, 2024).

Running these preliminary   experiments suggests that NickPay is well adapted to payments of sufficiently high values, e.g., international transfers or of financial assets of significant values, to offset their processing prices\footnote{More experiments and implementation analyses are warranted to see how these prices can be brought to lower values.}. Note that this is not a significant limitation, since the security properties that NickPay offers are of higher interest when the financial stakes are also higher. 

    \section{Related work}    
    \label{sec:related}

  We compare NickPay with existing related work, focusing on both privacy and auditability features in payment systems. 
  
  \subsection{Privacy}
    With ZeroCash, Sasson, Chiesa, Garman, Green, Miers \textit{et al}~\cite{sasson2014zerocash} propose a decentralized anonymous cash system  relying on Merkle trees and zero-knowledge proofs to anonymize
    the system and make it confidential,
    hiding the value of  transfers.
    It removes the bank authority and allows total anonymity for the user, i.e., no opening property is supported.


    Courtois and Mercer's approach~\cite{courtois2017stealth} enables anonymous transfers with minimal extra-communication and efficient computation.
    Indeed, after  registration, a sender can create a stealth address derived from a recipient public key without any help, i.e., with no external communication.
    The recipient has to scan payments over time and to try to unlock the stealth addresses with his secret key.
    This offers anonymity but no opening properties.
    It does, however, share similarities with NGS, such as deriving directly  anonymous recipient identifiers, and scanning to unlock transactions.
    
    Regarding existing stealth address schemes~\cite{courtois2017stealth}, Liu, Yang, Wong, Nguyen and Wang~\cite{liu2019key} pointed out that if a derived secret key, the secret key of a nickname in our case, is compromised, then all the secret keys of the corresponding user are compromised.
    Their approach insulates the secret keys to prevent this breach, thus offering an interesting key-insulation property.
    In our case, key insulation would facilitate malicious activities as it would be easy for a group member to grant an external user a derived key without compromising his own keys.
    Key insulation is not as desirable for NGS and NickPay as it is for stealth address schemes.

    \subsection{Auditability}

    Blazy, Canard, Fuchsbauer, Gouget, Sibert and Traoré's proposed in~\cite{blazy2011achieving} a transferable e-cash system, meaning a bank issues coins to users by signing them and users can transfer them to others anonymously; this system supports a judge entity capable of tracing coins and users.
    They use commuting signatures from randomizable Groth-Sahai proofs to ensure anonymity, and encryption for traceability, making the protocol heavy.
    Moreover, coins grow in size at every transaction, because their protocol adds details such as the recipient's public key at every interaction.

    Camenisch, Hohenberger and Lysyanskaya~\cite{camenisch2005compact} make e-cash solutions traceable for a cheating user. 
    Wüst, Kostiainen and Capkun~\cite{wust2019prcash} proposed audit capabilities when the predefined spending limit for a certain user is reached. NickPay addresses the auditability issue in its full extent.

    Narula, Casquez and Virza~\cite{narula2018zkledger} proposed a ledger supporting different audit queries. They used Pedersen commitments to hide the values in transactions between banks, and these banks can open up these commitments upon auditor's request. The auditor must therefore stay online to audit a bank, and supervision is not considered. A transaction requires heavy computation and verification, because its size is linear in the number of users.  Chatzigiannis and Baldimtsi~\cite{chatzigiannis2021miniledger} improved the space efficiency of this system, but the number of users is still limited due to the transaction size. 

    More recently, Yao and Zang~\cite{Yao22} used homomorphic encryption to preserve the privacy of transaction content, i.e., the  value and data assets, while an auditor is capable of decrypting transactions. For the privacy and auditability of transactions to be publicly verifiable, they require multiple proofs to be verified by the validation nodes; these proofs can be cumbersome. Moreover, they only focus on confidentiality, because anonymity can be handled by external privacy-preserving protection schemes. If  considered important, hiding transaction values could also be added to NickPay. 

    In 2022, Kiayias, Kohlweiss and Sarencheh proposed PEReDi~\cite{kiayias2022peredi}, a Central Bank Digital Currency (CBDC) offering interesting auditability capability and security properties. However, both the sender and receiver are required to complete transactions. 
    
    Finally, Sarencheh, Kiayias, and Kohlweiss~\cite{parscoin} introduce PARScoin, a stable-coin-based system that provides strong assurance in terms of privacy and auditability, in particular to check the constraints on off-chain funds induced by coin-stability requirements. Even though based on similar cryptographic technologies, NickPay, not being  linked to a particular currency, stable or not, uses a more general notion of auditability, better suited to implementing various banking regulations. 

    \section{Conclusion}
    NickPay is a  new privacy-preserving, auditable payment system built on top of the Ethereum permissionless blockchain platform, an issue of great interest to the financial world. It takes advantage of the recently introduced Nicknames for Group Signatures scheme for ensuring strong security properties regarding privacy and transaction independence. The demonstrated feasibility of an Ethereum-based NickPay prototype suggests that regulation-compliant payment systems could already be deployed as smart contracts on blockchains. 

    Future work could address possible improvements to NickPay performance or feature set. First, more experiments are needed to check NickPay's scalability. In particular, it could prove useful to introduce a ``tracing server''~\cite{van2013cryptonote}, e.g., at the supervisor level, to delegate the computation load that users bear to check all transactions to see if they can unlock those. 
    Secondly, the supervisor (and issuer) could be decentralized, as proposed in \cite{camenisch2020short}, with a threshold number of supervisors   to identify users; this would 
    avoid possible abusive surveillance. Thirdly, user revocation from groups must be addressed in detail. An obvious revocation strategy could  use a list of revoked members~\cite{boneh2004group}, although this would break some anonymity for past signatures of revoked users, which may not be acceptable. Finally, this specification of NickPay is account-based; one could envision other approaches, e.g., UTXO-based ones~\cite{nakamoto2008bitcoin}.

    \label{sec:conclusion}




    \bibliographystyle{acm}
\section*{Acknowledgment}
We thank BNP Paribas for supporting this research. 

    \bibliography{bib}

\end{document}